\documentstyle[11pt,adassconf]{article}  % Leave intact

\def\txitxo{Ben\'\i tez}
\def\etal{{et~al.}} 

\def\mysection#1{\vspace{-3pt}\section{#1}\vspace{-4pt}}
\def\align{{\tt align}}
\def\combDither{{\tt combDither}}
\def\combFilter{{\tt combFilter}}
\def\detectionCatalog{{\tt detectionCatalog}}
\def\filterCatalog{{\tt filterCatalog}}
\def\colorCatalog{{\tt colorCatalog}}
\def\pyblot{{\tt pyblot}}
\def\photoz{{\tt photoz}}

\begin{document}   % Leave intact
\tighten

%-----------------------------------------------------------------------
%			    Paper ID Code
%-----------------------------------------------------------------------
% Enter the proper paper identification code.  The ID code for your
% paper is the session number associated with your presentation as
% published in the official ADASS 2000 conference proceedings.  You can
% find this number locating your abstract in the printed proceedings
% that you received at the meeting or on-line via 
% http://adass2002.stsci.edu; the ID code is the letter/number sequence 
% proceeding the title of your presentation.  
%
% This will not appear in your paper; however, it allows different
% papers in the proceedings to cross-reference each other.
%
% EXAMPLE: \paperID{O4.1}
% EXAMPLE: \paperID{P7.7}
%
% Note that you should only have one \paperID, and it should not
% include a trailing period.  

\paperID{P7.8}

%-----------------------------------------------------------------------
%		            Paper Title 
%-----------------------------------------------------------------------
% Enter the title of the paper.
%
% EXAMPLE: \title{A Breakthrough in Astronomical Software Development}
% 

\title{An Automatic Image Reduction Pipeline for the Advanced Camera for Surveys}
%-----------------------------------------------------------------------
%		          Authors of Paper
%-----------------------------------------------------------------------
% Enter the authors followed by their affiliations.  The \author and
% \affil commands may appear multiple times as necessary (see example
% below).  List each author by giving the first name or initials first
% followed by the last name.  Authors with the same affiliations
% should grouped together. 
%
% EXAMPLE: \author{Raymond Plante, Doug Roberts, 
%                  R.\ M.\ Crutcher\altaffilmark{1}}
%          \affil{National Center for Supercomputing Applications, 
%                 University of Illinois Urbana-Champaign, Urbana, IL
%                 61801}
%          \author{Tom Troland}
%          \affil{University of Kentucky}
%
%          \altaffiltext{1}{Astronomy Department, UIUC}
%
% In this example, the first three authors, "Plante", "Roberts", and
% "Crutcher" are affiliated with "NCSA".  "Crutcher" has an alternate 
% affiliation with the "Astronomy Department".  The fourth author,
% "Troland", is affiliated with "University of Kentucky"
\author{John P.\  Blakeslee,	Kenneth R.\ Anderson,
        G.\ R.\ Meurer,	N.\ Ben\'\i tez}\vspace{-1pt}
%\affil{Department of Physics and Astronomy, Johns Hopkins University, 3400 N.\ Charles St., Baltimore, MD 21218}\vspace{-2pt}
\affil{Physics \& Astronomy, Johns Hopkins University, Baltimore, MD 21218}\vspace{-2pt}
\author{D.\ Magee}\vspace{-1pt}
\affil{UCO/Lick Observatory, University of California, Santa Cruz, CA 95064}

%-----------------------------------------------------------------------
%			 Contact Information
%-----------------------------------------------------------------------
% This information will not appear in the paper but will be used by
% the editors in case you need to be contacted concerning your
% submission.  Enter your name as the contact along with your email
% address.
% 
% EXAMPLE:  \contact{Dennis Crabtree}
%           \email{crabtree@cfht.hawaii.edu}
%
\contact{John Blakeslee}
\email{jpb@pha.jhu.edu}

%-----------------------------------------------------------------------
%		      Author Index Specification
%-----------------------------------------------------------------------
% Specify how each author name should appear in the author index.  The 
% \paindex{ } should be used to indicate the primary author, and the
% \aindex for all other co-authors.  You MUST use the following
% syntax: 
%
% SYNTAX:  \aindex{LASTNAME, F. M.}
% 
% where F is the first initial and M is the second initial (if
% used).  This guarantees that authors that appear in multiple papers
% will appear only once in the author index.  
%
% EXAMPLE: \paindex{Crabtree, D.}
%          \aindex{Manset, N.}        
%          \aindex{Veillet, C.}        
\paindex{Blakeslee, J. P.}
\aindex{Anderson, K. R.}     % Remove this line if there is only one author
\aindex{Meurer, G. R.}     % Remove this line if there is only one author
\aindex{Benitez, N.}     % Remove this line if there is only one author
\aindex{Magee, D.}     % Remove this line if there is only one author
%-----------------------------------------------------------------------
%			Subject Index keywords
%-----------------------------------------------------------------------
% Enter up to 6 keywords describing your paper.  These will NOT be
% printed as part of your paper; however, they will be used to
% generate the subject index for the proceedings.  There is no
% standard list; however, you can consult the indices for past ADASS
% proceedings (http://iraf.noao.edu/ADASS/adass.html). 
%
% EXAMPLE:  \keywords{visualization, astronomy: radio, parallel
%                     computing, AIPS++, Galactic Center}
%
% In this example, the author noticed that "radio astronomy" appeared
% in the ADASS VII Index as "astronomy" being the major keyword and
% "radio" as the minor keyword.

%-----------------------------------------------------------------------
%			       Abstract
%-----------------------------------------------------------------------
% Type abstract in the space below.  Consult the User Guide and Latex
% Information file for a list of supported macros (e.g. for typesetting 
% special symbols). Do not leave a blank line between \begin{abstract} 
% and the start of your text.
\vspace{-2pt}
\begin{abstract}          % Leave intact
We have written an automatic image processing pipeline for the
Advanced Camera for Surveys (ACS) Guaranteed Time Observation
(GTO) program.  The pipeline, known as Apsis,
supports the different cameras available on
the ACS instrument and is written in Python with a flexible 
object-oriented design that simplifies the
incorporation of new pipeline modules.  
% It makes use of the PyFits and Pyraf packages distributed by STScI,
% as well as other external software.  
The processing steps include empirical
determination of image offsets and rotation, cosmic ray rejection,
image combination using the drizzle routine called via the STScI
Pyraf package, 
object detection and photometry using SExtractor, and photometric
redshift estimation in the event of multiple bandpasses.
%The products are converted to XML
The products are encapsulated in XML markup
for automated ingestion into the ACS Team archive.
\end{abstract}
%\vspace{-12pt}
\vspace{-6pt}

\section{Introduction and Basic Operation}

The ACS science team was apportioned about 550 orbits of GTO time in
exchange for its instrument development and calibration work.
A robust and flexible astronomical data reduction and analysis pipeline 
was required for processing these data and other observations 
for which the science team is responsible.
% that fall under the purview of the ACS science team.
%%
%Nature abhors a vacuum, so
To fill this need, we developed a package called ``Apsis'' (ACS
pipeline science investigation software), written in the Python 
%%programming language using a highly flexible, modular design.
programming language using a flexible, modular design.
% to simplify the incorporation of new processing steps and allow
% the omission of some steps for certain types of dataset. 
Apsis was used for processing 
the ACS early release observation (ERO) images of the Tadpole and Mice
galaxies, the Cone Nebula, and M17 
%%soon after the installation of ACS on the Hubble Space Telescope. 
soon after ACS was installed on the Hubble Space Telescope. 
Since then, it has developed and been used extensively 
on Linux and Solaris platforms for processing WFC and HRC
data from both science and calibration programs.
% and active development work continues.
%Apsis development is ongoing.

% Apsis is written in the Python programming language using an
% object-oriented, modular design to simplify the incorporation of new
% processing steps and allow the omission of some steps for certain types
% of dataset.

A given processing run starts with flat-fielded multi-extension FITS images
that have been processed through the CALACS software  at STScI (Hack 1999). 
The basic Apsis ``observation object'' consists of all the
images of a given program field (or mosaic of adjacent fields). 
These are grouped into separate
``associations'' according to the different imaging/grism filters
used or (sometimes) different epochs, and FITS tables are 
constructed similar to those used in STScI OPUS pipeline.
%% does this fit here?
Reading and manipulation of FITS images and tables are done
with the aid of the PyFits and Numarray Python modules,
and any IRAF/STSDAS routines are called through Pyraf
(see Greenfield \etal\ 2002).
Other external programs are accessed via system calls.

The observation object is then passed to the various python modules
in succession, which have ``methods'' (python functions) for
performing the following general tasks:
image offset measurement corrected for distortion,
sky estimation and subtraction, cosmic ray (CR) rejection and ``drizzling,''
construction of error arrays and a multi-band ``detection image,''
object detection and measurement, photometric calibration,
and photo-$z$ estimation.
A running log keeps track of progress and records diagnostics.
After completion, each module 
object is written to disk as a byte stream using the Python ``pickling''
utility; this allows for recovery of information in the event
of failure and simplifies debugging.  The modules also contain
% sophisticated detailed elaborate complex florid embroidered
detailed methods for producing XML messages and markup of the data
products (images and catalogs) for archiving purposes. 

Although Apsis was designed primarily as an automated pipeline,
it can be run by users as a standalone program, and provides a
suite of command-line options for this purpose.  These include
switches for turning off various parts of the processing (e.g., 
sky subtraction, cataloging), specifying a particular input
image as reference for the shifts, supplying an alternate distortion
model, externally measured shifts, or sky values, lowering
the CR rejection thresholds, changing the output image
pixels scale, and several other options.  The following sections
provide a few details on the major processing steps.

\mysection{Image Registration and Sky Subtraction} 

The \align\ module  determines the relative $x,y$ shifts and rotations,
as well as the sky levels, of the input images.
SExtractor (Bertin \& Arnout 1996) is run with a signal-to-noise threshold of 10
on each science extension of each image
%(1 for the HRC; 2 for the WFC) 
(there are two science extensions for WFC; one for HRC and SBC).
% The resultant catalogs are culled for ``good'' sources: those not likely to
% be cosmic rays, CCD artifacts, or overly diffuse objects based on full-width
% at half maximum (FWHM), rms width along the minor axis, and ellipticity.
The resultant catalogs are culled on the basis of object
size and shape parameters, thereby rejecting the vast majority of cosmic 
rays and CCD artifacts as well as overly diffuse objects.
If fewer than ten ``good'' sources remain, then SExtractor
is rerun at lower thresholds.  The $x,y$ coordinates of each ``good'' source 
are corrected using the distortion model read from the IDCTAB FITS table
specified in the image headers or on the command line.  Sources from
different extensions (different CCD chips) of the same image are placed
on a common rectified frame 
%using the V2REF, V3REF IDCTAB parameters.
using the IDCTAB parameters V2REF and V3REF.

We use the ``Match'' program (Richmond 2002) to derive shifts and
rotations with respect to a reference image, by default the 
one having the most ``good'' sources.   We modified Match to accept 
an input guessed transformation (derived from the headers)
and to report more diagnostics for evaluating the success of
the matching.   In the event of failure, Match
is rerun without an input guess, using the full triangle-matching
search algorithm on the $N$ brightest sources; this is repeated with
larger $N$, and an $N^6$ hit in processor time,
if necessary.  All sources are used for tuning up the final 
transformation once it is found.
We also evaluate the median $x,y$ shifts for all matched sources,
and we revert to these if the derived rotation is negligible.

This automatic, adaptive matching procedure works quite well in practice.  
Images taken at large offsets and in different visits can have header shift
errors of ${\sim\,}1\arcsec$ (${\sim\,}$20 WFC pixels), but this is irrelevant for the
triangle matching algorithm.  Typical WFC GTO fields produce
one-to-several hundred matched sources per exposure, and the
resulting shift uncertainty is typically ${\sim\,}0.02$ pix.
Problems have only occurred for some images of blank fields
taken with the HRC, which has an area 1/64 of the WFC.
In this case Apsis defaults to the header shifts,
although it is also possible to supply external shifts.
Grism (G800L) images are by default aligned with direct images
according to the headers.

The sky values and sigmas returned by SExtractor for each image
extension are used as inputs to the STSDAS dither.sky task,
which in turn uses the STSDAS gstatistics routine.
However, experimentation indicated that the mean sky levels reported
by SExtractor were more robust, although tended to be biased high
in more crowded fields.  For this reason, we adopt the SExtractor
sky value when it is the lower estimate, and the mean of the two
otherwise.  The sky is then subtracted, with the option
of averaging the values for different extensions of the same image.
This step removes sky level differences for the image combination and
CR rejection routines described in the following section.

\mysection{Image Combination}

The modules \combDither\ and \pyblot\ combine all exposures
within each filter into a single geometrically corrected image while
rejecting cosmic rays.  This is done through the STSDAS Dither
package drizzle-blot-drizzle cycle outlined by Gonzaga \etal\ (1998), although
here coded in Python.   First the images are ``drizzled'' to 
separate output images using the shifts and rotations supplied
by the Apsis \align\ module.  These individually drizzled images are 
then median stacked using exposure time weighting and a
`minmax' clipping algorithm.
% and rejection parameters depending on the number of input images.
The median image is then ``blotted''
back to each of the input image positions, rescaled in
exposure time, and used as a template for CR rejection
with the driz\_cr task.  The driz\_cr parameters were optimized
to achieve good CR rejection without harming the centers of any stars.
In particular, this meant setting the derivative scale parameter
to a value near unity.  

A cosmic ray mask is produced for each science extension and is multiplied
by the bad pixel mask that we produce from the data quality arrays using a
call to Pydrizzle  with the `bits' parameter set to 8578
to include `good', `replaced', `saturated,' and `repaired' pixels
(see Hack 1999).  These combined CR/badpix masks are then used
for the final drizzling of the input images in each filter to a single
output image.  We again use the shifts and rotations found by \align\
and produce drizzled images in units of electrons.  In so doing,
it is necessary to divide the output pixel values by the number of science
extensions, since the drizzle task does not recognize the different
extensions as being part of the same image.  The default drizzle
pixfrac and scale parameters are unity but can be reset with flags
on the Apsis command line.

After the final images are produced, the image headers are thoroughly
updated and corrected.  This includes calculating new CD matrices from scratch
based on the output pixel scale and image orientation, which is derived
through spherical geometry from the PA\_V3 header keyword of the reference
image, the declination, and the detector orientation and
location in the V2,V3 plane.  The code also ensures that all of the output
images have identical WCS information, as they should all be well
aligned at this stage.  At this writing, the WCS zero points are
off by roughly 2$\arcsec$ due to systematic pointing errors, 
but this is expected to improve with a recent FGS realignment.

We produce an ``RMS image'' for each output
science image.  The RMS images have pixel values equal to the estimated
error per pixel, based on the total read noise, signal level, Apsis cosmic
ray rejection, and the effects of the multiple bias and
dark frame subtractions.  The actual root-mean-square pixel variation in
the science image is of course lower due to the noise correlation induced
by non-integer shifts and geometric correction.  However, we verified
that the RMS images reflect the pixel variation when images are stacked
without shifting or correction.  The RMS images are used 
% for the object photometry.
in estimating photometric errors.
%to estimate the errors in the object photometry.

\mysection{Object Detection, Photometry, and Redshift Estimation}

A series of modules do the object photometry.
The \combFilter\ module produces a multi-band ``detection image,''
which is the variance-weighted average of the science images in the 
different bandpasses.  Images in specific filters (e.g., grism, polarizers)
can be omitted from this average.  This module also produces a 
detection weight image, which is a similarly weighted sum of the exposure
maps from the drizzling process.  The \detectionCatalog\ and
\filterCatalog\  modules create parameter files and run SExtractor, first on the
detection image alone, and then multiple times in ``dual image mode.''
%with on-the-fly editing of the input parameters.
In dual mode, the detection image defines the object position and apertures, 
but the photometric measurements are done on the filter image
with its associated RMS image.

The \colorCatalog\ module derives the photometric zero points from
the image header catalogs, sorts through the different SExtractor
catalogs, and produces a single, calibrated multicolor catalog.  
It also uses simple logic for
flagging apparently  bad or anomalous magnitudes.  Finally, the
\photoz\ module feeds the multicolor catalog to the Bayesian
Photometric Redshift package (\txitxo\ 2000), which is 
itself coded in Python.  A final XML ``run message''  is written
in the same format as the individual module messages.  
The images and XML catalogs can
then be ingested into the ACS team data archive.

\mysection{Summary and Outlook}

We have written a Python package called Apsis which provides a highly
robust, efficient, and self-documenting means of reducing ACS GTO data.
Apsis was used for processing the ERO datasets
released to the press soon after launch.
Several improvements are still envisioned, including
improved object detection for crowded fields,
higher order resolution-preserving drizzle kernels,
and on-the-fly calibration of the WCS zero points using
online astrometric catalogs.

\par\vskip 9pt \noindent
%\acknowledgments
%%ACS was developed under NASA contract \hbox{NAS\,5-32864}, and 
Apsis development was supported by NASA grant NAG5-7697.  
We thank our many ACS team colleagues 
% who helped facilitate the development of this software.
% who provided help, advice, and feedback for this work.
for help, advice, and feedback.
We also thank Colin Cox, Warren Hack, and Richard Hook for helpful discussions.
\end{document}